\documentclass[12pt]{article}
\usepackage{graphicx}
\usepackage{amssymb,amsmath,amsfonts,palatino,amsthm}
\usepackage{amssymb}
\usepackage{epstopdf}
\usepackage{slashed} 
\newcommand{\f}{\begin{equation}}
\newcommand{\ff}{\end{equation}}
\newcommand{\blankline}{\vskip .3cm}
\begin{document}

\title{Quantum mechanics and the principle of maximal variety \\}
\author{Lee Smolin\thanks{lsmolin@perimeterinstitute.ca} 
\\
\\
Perimeter Institute for Theoretical Physics,\\
31 Caroline Street North, Waterloo, Ontario N2J 2Y5, Canada}
\date{\today}
\maketitle

\begin{abstract}

Quantum mechanics is derived from the principle that the universe contain as much variety as possible, in the sense of maximizing the distinctiveness of each subsystem.

The quantum state of a microscopic system is defined to correspond to an ensemble of subsystems of the universe with identical constituents and similar preparations and environments.   A new kind of interaction is posited amongst such similar subsystems which acts to increase their distinctiveness, by extremizing the variety.  In the limit of large numbers of similar subsystems this interaction is shown to give rise to Bohm's quantum potential.  As a result the probability distribution for the ensemble is governed by the Schroedinger equation.

The measurement problem is naturally and simply solved.  Microscopic systems appear statistical because they are members of large ensembles of similar systems which interact non-locally.  Macroscopic systems are unique, and are not members of any ensembles of similar systems.  Consequently their collective coordinates may evolve deterministically.  

This proposal could be tested by  constructing quantum devices from entangled states of a modest number of quits which, by its combinatorial complexity, can be expected to have no natural copies.

\end{abstract}

\newpage

\tableofcontents


\section{Introduction}
  
This paper presents a new completion of quantum mechanics based on three key ideas.  The first is that 
quantum mechanics is necessarily a description of subsystems of the universe.  It is an approximation to some other, very different theory, which might be applied to the universe as a whole\cite{SU,TR,TN}.

The second idea is that the quantum state refers to an ensemble of similar systems present in the universe at a given time.  We call this the {\it real ensemble hypothesis\cite{RE}.}  By similar systems we mean systems with the same constituents,  whose dynamics are subject to (within errors that can be ignored) the same Hamiltonian, and which have very similar histories and hence, in operational terms, the same preparation.  The very peculiar idea underlying this proposal is that such similar systems have a new kind of interaction with each other, just by virtue of their similarities. This interaction takes place amongst similar systems, regardless of how far apart they may be situated in space, and thus, if these ideas turn out to be right,  is how non-locality enters quantum phenomena. 

This hypothesis is motivated by a line of thought involving the application to quantum gravity and quantum mechanics of some very general principles.  But in the interest of getting quickly to the point, this motivation is postponed untill section 6.  We will only say here that quantum gravity points to the possibility that space and locality are both emergent and that, if this is so, we should expect there to be defects in locality, where events are connected which are far separated in the emergent low energy classical metric.   

If the reader will take this {\it real ensemble} hypothesis as a provisional idea, he or she will see that, together with the third idea, it leads to quantum mechanics.  

 The real ensemble hypothesis was explored earlier in \cite{RE}, where a new kind of interaction amongst the similar systems which make up the ensemble was posited and shown to yield quantum dynamics.  However that work could be criticized because the particular inter-ensemble interaction was complicated and motivated only by the fact that it gave the right answer.  

A similar proposal was made in \cite{MIW,DHW}, where the idea of {\it many interacting classical worlds (MIW)} was introduced to explain quantum mechanics.  This also posits that the quantum state refers to an ensemble of real, existing, systems which interact with each other, only those were posited to be near copies of our universe that all simultaneously exist\footnote{If the ontology posited by the \cite{MIW,DHW} papers may seem extravagant, their proposal had the virtue of a simple form for the inter-ensemble interactions.  This inspired me to seek to use such a simple dynamics in the real ensemble idea.  In particular, an important insight contained in \cite{DHW} is that if there are $N$ particles on a line with positions, $x_i$, with $i =1, \dots, N$, the density at the $k$'th point can be approximated by
\f
\rho (x_k ) \approx \frac{1}{N (x_{k+1} - x_k )}
\ff
This motivates the choose of a ultraviolet cutoff, in equation (\ref{cutoff}) below.}.

The third key idea is that the inter-ensemble interaction can be related to the an observable called the {\it variety} of the collection of similar subsystems.  This is also motivated by the general principles, as will be explained in section 6.

{\it The principle of maximal variety}, was formulated with Julian Barbour in the 1980's\cite{variety}.  The variety of a system of relations, $\cal V$, is a measure of how easy it is to distinguish the neighbourhood of every element from that of every other.   

The basic idea can be applied to every system with elements $e_i$, labeled by $i = 1, \ldots , N$ whose dynamics depends on relational observables,  $X_{ij}$.  The elements could be particles or events or subsystems and the relations could be relative position, relative distance, causal relations, etc.  We proceed by defining the {\it view} of the $i$'th element, this summarizes what element $i$ may ``know" about the rest of the system by means of the relational observables.  The view of $i$ is then denoted $V_i (X_{ij} )$.

Different systems will be described by different views.  For example, if the system is $N$ points in a $d$ dimensional Euclidean space, the view of the $k$'th point is the list of vectors to the other points, weighed by the distance. 
\f
V_i^{k a} = \frac{x_i^a - x_k^a }{D(i,k)^2} = \frac{x_i^a - x_k^a }{|x_i^a - x_k^a |^2}
\ff
We then define the {\it distinctiveness} of two elements, $i$ and $j$ to be a measure of the differences between the views of $i$ and $j$.  If the views live in a vector space this can be denoted,
\f
{\cal I }_{ij} = | V_i -V_j |^2
\ff
 The variety is then defined to measure the distinguishability of all the elements from each other.
\f
{\cal V} = \frac{1}{N(N-1)} \sum_{i \neq j} {\cal I}(i,j)
\ff

The distinctiveness provides a metric on the set of subsystems.
\f
h_{ij} = {\cal I }_{ij} = | V_i -V_j |^2
\ff
This metric tells us that systems are close if they have similar views of their relations to the rest of the universe.
For example,  two events that are close in space will have similar views.  

But this is not the only circumstance in which two events may have similar views.  If the events are each part of the history of a microscopic system, that can each be considered to be effectively isolated, and if those two isolated systems have similar constituents, environments and preparations, than their views may also be similar.  
  
Now assuming conventional notion of locality, the degrees of freedom at nearby events can interact.  But suppose locality in space is not primary.  Suppose, instead, that the metric on the space of views is actually what determines the relevant notion of locality for interactions.  As a consequence, two systems may interact when they are nearby in space, {\it or when their views are similar because they have the similar constituents, environments and preparations.}  The former give conventional local interactions, while the latter case gives a new kind of interactions.  The aim of this paper is to show that the latter kinds of interactions may be responsible for quantum phenomena.

In particular, we will show that when these new interactions amongst members of the ensemble of systems with similar views acts to increase the variety of the beables in that ensemble, they give rise to the quantum potential of Bohmian quantum mechanics.  This can be made plausible if we consider the fact that the Bohmian potential is repulsive and so acts to smooth out the wave function, giving rise to a greater variety of beables represented in the ensemble.  

There are then three basic hypotheses in this paper.

\begin{enumerate}

\item{}In the microscopic causal geometry underlying nature, two systems can interact if they are within a distance $R$ in the metric $h_{ij}$.  There are two ways this can happen.  It can happen when they are nearby in the emergent macroscopic notion of spatial geometry.  When two people stand next to each other and scan a landscape they see similar views.  But two microscopic systems can also be very far apart in the macroscopic geometry and still have a similar view of their surroundings.  When this happens there are a new kind of interactions between them.

\item{}Similar systems, nearby in $h_{ij}$ but distant in space, form ensembles that mutually interact.  It is these ensembles that the quantum state refers to.  

\item{}These new interactions are defined in terms of a potential energy which is proportional to the negative of the variety.  Lower energy implies higher variety.

\end{enumerate}

We will see that these new interactions between members of the ensembles that define quantum states give rise to quantum phenomena. 

\blankline
\blankline

The main result of this paper is that when $N$, the number of subsystems in the ensemble, is large, the variety can be expressed in terms of a probability density for the ensemble and that, when so expressed, $\cal V$ is closely related to the quantum potential of Bohm.  Consequently the evolution of the ensemble probabilities is given by the Schroedinger equation.  A second result is a prediction of specific corrections that arise and are expressed as non-linear  corrections to the Schroedinger equation.  

Another result is that the measurement problem is naturally and simply solved.  Microscopic systems appear statistical, when described as local systems in isolation,  because they are members of large ensembles of similar systems which with they interact non-locally.  Macroscopic systems are unique, and are not members of any ensembles of similar systems. Not being members of large ensembles, their collective coordinates are not disturbed by non-local interactions with distant subsystems, nor can they be described by quantum states.   Consequently their collective coordinates may evolve deterministically. 

The precise proposal for a non-local completion of quantum mechanics is presented in section 2, while section 3 is devoted to deriving quantum mechanics as an approximation in the limit that $N$, the number of subsystems in the ensemble goes to infinity.  In section 4 we discuss several experimental tests that become possible in cases where $N$ is small, while section 5 considers several possible objections to these results.  Then finally, in section 6, we discuss in more detail the motivation for this proposal\footnote{Some possibly related approaches are \cite{entropic}.}.

\section{The dynamics of extremal variety}

We now begin the formal development, by which we derive quantum mechanics from the principles we have described.

We consider an ensemble of $N$ identical systems, each of which lives in a configuration space we will for simplicity take to be $R^d$, coordinatized by $x^a_k$, where $a=1, \ldots d$ and $k=1, \ldots , N$.  

All the relational information about a subsystem of the universe is contained in the {\it view} that subsystem has of the rest of the universe, through its causal links or other relations to other subsystems.  This is the central element we will employ in our reconstruction of quantum theory.

Let us then define the {\it  view of the $i$'th system},
\f
\boxed{
V_i^{k a} = \frac{x_i^a - x_k^a }{D(i,k)^2} = \frac{x_i^a - x_k^a }{|x_i^a - x_k^a |^2} \Theta (R-|x_i^a - x_k^a | )}
\ff
for $k \neq i$,
be seen, for each $i$, as a vector of components labeled by $k$, each component of which is a vector, that shows the system $i$'s relations to its $N-1$ neighbours.   The vector's  components are weighed by the distance.
This can be called the {\it view} of the rest of the system, experienced by the $i$'th element.  The closer $k$ is to $i$, the larger is $V_i^k$ and the more important $k$ is to $i$'th view of the rest of the system.

Note that we insert a cutoff $R$ on the view, by choosing the distance function to be 
\f
D(i,k)^2 = |x_i^a - x_k^a |^2 \frac{1}{\Theta (R-|x_i^a - x_k^a | )}
\label{Rcutoff}
\ff
Thus if the $k$'th system is more than a metric distance $R$ from the $i$'th system it ``falls outside the horizon" and is an infinite distance away.  The  cutoff $R$ will play a role in what follows.


Now let us construct a measure of the differences between two elements $i$ and $j$.
We can simply take the difference of the two vectors, $V_i^k $ and $V_j^k$.
\f
{\cal I}_{ij} = \frac{1}{N} \sum_k \left ( V_i^{k a} - V_j^{k a} \right )^2   
\ff
can be called the {\it distinctiveness of $i$ and $j$.}  The larger ${\cal I}_{ij}$ are the more easily they can be differentiated by their views.

To get the variety we sum this over all the pairs $i  \neq j$
\f
{\cal V} =  \frac{A}{N^2} \sum_{i \neq j}  {\cal I}_{ij} =  \frac{A}{N^3} \sum_{i \neq j} \sum_k \left ( V_i^k - V_j^k \right )^2   
\ff
where $A$ is a dimensionless normalization constant.

So we  define a new inter-ensemble potential energy as 
\f
{\cal U}^{\cal V} =- \frac{ \hbar^2}{8m}{\cal V}
\label{UV}
\ff 
We choose to posit that the potential energy is the negative of the variety so that in the ground state the variety will be maximized.
Note that $\cal V$ has dimensions of inverse-length-squared and that this potential is negative definite.   The constant
$\frac{\hbar^2}{m}$ is necessary for dimensional considerations, to turn an inverse area into an energy.  Of course an $\hbar$ is required if we are to make good on our claim that these new interactions give rise to quantum phenomena.


We can think of $\cal V$ in a different way, as a local function in $x_k$, by reversing the order of the summations.
\f
{\cal V} = \frac{A}{N} \sum_{z}  {\cal V}_{z} =  \frac{A}{N^3} \sum_k\sum_{i \neq j}  \left ( V_i^{k a } - V_j^{k a } \right )^2   
\ff
where
\f
{\cal V}_{k} =  \frac{1}{N^2} \sum_{i \neq k \neq  j}  \left ( V_i^{k a } - V_j^{k a } \right )^2
\ff

\subsection{The fundamental dynamics}

To write a dynamical theory we need to introduce momenta beables $p_{a}^i$, in addition to  the position beables $x^a_{ i}$.  

However the correspondence with quantum mechanics requires that in the large $N$, continuum limit, the momenta of the particles, $p_{a}^i$, merge into a momentum density $p_{a} (x)$. This, moreover, must be a gradient of a phase 
$S(z)$, which comes from the decomposition of the wave function.  
\f
p_a (x^a ) = \partial_a S (x^a)   
\ff
where $S$ is related to a complex phase, $w (x^a )= e^{\frac{\imath}{\hbar}S}$  subject to
\f
w^* (x^a ) w (x^a ) =1
\ff 

Consequently,  Takabayashi and Wallstrom\cite{wallstrom,T1952} noted that further conditions are needed to guarantee that $e^{\imath S/\hbar }$ is single valued.  We can address this by replacing the momenta beables, $ p_a^i$ with a complex phase factor beable (one for earh subsystem), associated,
\f
w_i ,  \ \ \ \  w^*_i w_i =1
\ff 
We can write 
\f
w_k = e^{\frac{\imath}{\hbar}S_k}
\label{Sdef}
\ff
but remember that $S_k$ is only defined modulo $2 \pi \hbar$.

Mindful that we want to express the theory symmetrically in all pairwise relationships, we will posit that the momenta $p_a^k$ are composite variables which code a subsystem's view of the ratios of the phase factor beables.
\f
p_{k a} = - \imath \frac{1}{N} \sum_{j \neq k} V_k^{a j}  \ln{ \left ( \frac{w_{j}}{w_k} \right ) } 
\ff
So we write the kinetic energy as
\f
K.E. = {\cal R}e \frac{Z \hbar^2 }{2m N^2 }\sum_{k \neq j} \frac{1}{(x_k -x_j )^2 } 
\left [
\ln{ \left ( \frac{w_{j}}{w_k} \right ) } 
\right ]^2 
\label{KE}
\ff
where $Z$ is a normalization constant to be determined.  

Putting this together with the inter-ensemble potential energy we have the fundamental action,
\f
S(w,x)  =  \int dt \sum_k \left \{  -  Z_0  \sum_{j \neq k} 
 {x}_k^a \imath V^{j a}_k \frac{d}{dt}  \left [  \ln{ \left ( \frac{w_{j}}{w_k} \right ) } \right ]   -
H[x,w] \right \}
\ff
where 
\f
\boxed{ H[x,w] =   \frac{ \hbar^2 }{2m} \left (   Z  \sum_{k \neq j} ( V_k^{j a} )^2
\left [  {\cal R}e
\ln{ \left ( \frac{w_{j}}{w_k} \right ) } 
\right ]^2 
-\frac{ A}{4N } \sum_k\sum_{i \neq j}  \left ( V_i^{k a } - V_j^{k a } \right )^2  \right )
 + \sum_{k }  U (x_k ) }
\ff
Here $Z,Z_0$, and $A$ are normalization constants and $U(x) $ is an ordinary potential energy.

Note that we have used the relative locality form for the symplectic potential\cite{rl}.  
\f
S_0 = - Z_0 \int dt \sum_k   \sum_{j \neq k} 
 {x}_k^a  \dot{p}_a^k  
 = -  Z_0  \int dt \sum_k  \sum_{j \neq k} 
 {x}_k^a \imath V^{j a}_k \frac{d}{dt}  \left [  \ln{ \left ( \frac{w_{j}}{w_k} \right ) } \right ]  
\ff

Our task is now to show that when $N$ is large this is equivalent to ordinary quantum mechanics.


\section{Derivation of quantum mechanics}

We first evaluate  the inter-ensemble potential energy, then we do the same for the average of the kinetic energy and the symplectic potential.

\subsection{The origin of the quantum potential}

To express $<{\cal V} >$ as an integral over local functions we write, for a function $\phi (x )$, 
\f
< \phi > = \frac{1}{N} \sum_k \phi (x_k )  \rightarrow \int  d^d z \rho (z) \phi (z)
\label{basic}
\ff
In the limit $N \rightarrow \infty$ this defines the probability density for configurations, $\rho (z)$.

Similarly we turn the sums on $i$ to an integral,
\f
\frac{1}{N} \sum_i \phi (x_{k+i}, x_k )  \rightarrow Z \int_a^R  d^d x \rho (z+x) \phi (z+x,z)
\ff
and similarly for $j$.  The possibility that the integral only approximates the sum for finite $N$, because of the roughness of the estimate for the limits on the integral,  is accounted for by an adjustable normalization factor $Z$.

Note that  we have to be careful to impose limits on the integral to avoid unphysical divergences in $  \frac{1}{x}$.  
These divergences are unphysical because for finite $N$ two configuration variables, $x^a_k$ and $x_j^a$, cannot come closer than a limit which varies inversely with the density at $x^a_k$ and $N$.  This is because if $x_k^a$ and $x_j^a$ are nearest neighbours in the distribution, the density at one of their locations is related to their separation.
\f
\rho ( x_k^a ) \approx \frac{1}{N |x_k - x_j |^d }
\ff
Hence, for finite $N$ they are very unlikely to coincide. When we approximate the sums by integrals, the integrals representing intervals between configurations must then be cut off by a short distance cutoff $a$ that scales inversely like a power of $N \rho (z)$.
The short distance cutoff $a (z)$ on the integral above in $d^d x$ then expresses this fact that there is a limit to 
$ \frac{1}{x}$  related to the density.   Hence the short distance cutoff is at
\f
a(z) =\frac{1}{( N \rho (z))^{\frac{1}{d}}}
\label{cutoff}
\ff
There is also an infrared cutoff, $R$ coming from (\ref{Rcutoff}). This tells us that two systems further than $R$ in configuration space do not figure in each other's views.   A key question turns out to be how the physical cutoff scales
with $N$.  We will define
\f
r'= N^{\frac{1}{d}}R
\ff
to represent a fixed physical lengths scale which is held fixed when we take the limit of large $N$ at the end of the calculation.  That way, the physical ultraviolet and infrared cutoffs scale the same way with $N$.  But the large scale, infrared cutoff, $r^\prime$ can't know about the value of the probability distribution at some far off point $z$, so while $a$ scales with $\rho$, $r^\prime$ doesn't.

As a result when we scale $x$ and $d^d x$ with $a$ to make the integrals dimensionless, we define $r$, such that,
$R=ar$. But we then hold fixed 
$r'=\frac{1}{\rho^{\frac{1}{d}}}r= N^{\frac{1}{d}}R$ as we take $N$ large.  
$r^\prime$, unlike $r$, is a length. We shall see that $r^\prime$  defines a new physical length scale at which the linearity of quantum mechanics gives way to a non-linear theory.

Thus, the continuum approximation to the variety  is,
\f
 {\cal V}  =   \int d^d z \rho (z) Z_V  \int_a^R  d^d x \int_a^R  d^d y
 [ (  
\frac{x^a}{x^2 } - \frac{y^a }{y^2}  )^2   \rho (z+x ) \rho (z+ y ) 
\ff

We do a scale transformation by writing
$x^a =a \alpha^a $ and $y^a =a \beta^a $.  
To get a single integral over a local function we can expand
\f
\rho (z+a \alpha) = \rho (z) + a \alpha^a  \partial_a \rho (z) + \frac{1}{2} a^{ 2} \alpha^a \alpha^b \partial^2_{ab} \rho (z) +  \ldots
\ff
and similarly for $\rho (z+ a \beta) $ and perform the integrations, holding the upper limit $r^\prime$ fixed. 

The  normalization factor  is
\f
Z_V=\frac{1}{N^3} \frac{d^2 N^2}{\Omega^2 (r^d-1)^2 } \approx  \frac{d^2}{2N \Omega^2 r^{2d} }
\ff
The result is
\f
{\cal V} =   \int d^d z \rho \left (  \frac{1}{R^2 } - 
 ( \frac{1}{\rho } \partial \rho )^2 +\frac{1}{N^{\frac{2}{d}}} \frac{d}{d+2} r^{\prime 2} \frac{(\nabla^2 \rho )^2}{\rho^2}
+\ldots
\right )
\ff
Here we ignore total derivatives, which don't contribute to the potential energy.  
The first term is an ignorable constant.  The second term is what we want;  its variation gives the Bohmian quantum potential.  


The higher order terms are suppressed by powers of $\frac{1}{N^{\frac{2}{d}}}$.  The result is 
\f
{\cal U}^{\cal V} = -\frac{ \hbar^2}{8m}{\cal V} =
\frac{ \hbar^2}{8m} \int d^d z \rho (\frac{1}{\rho} \partial_a \rho )^2 +    O(\frac{1}{N^{\frac{2}{d}}} )
\ff 
which we recognize as the term whose variation gives the Bohmian quantum potential. 

The leading correction is 
\f
{\cal U}^{\Delta {\cal V}} = -\frac{ \hbar^2}{8m}\Delta {\cal V} =
- \frac{1}{N^{\frac{2}{d}}} \frac{ \hbar^2 r^{\prime 2}}{8m} \int d^d z \rho (\frac{1}{\rho} \nabla^2  \rho )^2 
\ff
which contributes non-linear corrections to the Schroedinger equation.

\subsection{The kinetic energy}

We can similarly evaluate the kinetic energy.
We write the continuum approximation, using a function $w (x)$ defined so that
\f
w (x_k ) =w_k
\ff
This is possible because the inter-ensemble interaction is repulsive so it would require infinite potential energy for two configurations to sit on top of each other.  So there are never two members of the ensemble $k$ and $j$ such that $x_k^a =x_j^a$.  Thus, if there is a member of the ensemble sitting at a point $x^a$ then it is unique and we can assign a definite $w (x_k ) = w_k$ to it.  

We find using (\ref{KE},\ref{basic})
\begin{eqnarray}
K.E.&=&  {\cal R}e  \frac{ \hbar^2  }{2m} \int d^d z \rho (z) Z_{KE} \int_a^R d^d  x \rho (x) 
\frac{1}{(z -x )^2 } 
\left [
\ln{ \left ( \frac{w (x)}{w (z)} \right ) } 
\right ]^2 
\\
&=& \frac{1   }{2m} \int d^d z \rho (z) \int_a^R d^d  x \rho (x) Z_{KE}
\frac{1}{(z -x )^2 } 
\left [ S (x) -S(z)   
\right ]^2 
\end{eqnarray}
where we recall that, $ w (x) = e^{\frac{\imath}{\hbar}S(x)}$.

We rewrite in terms of $a\alpha = x-z$ and expand in powers of 
$a = \frac{1}{(N\rho )^{\frac{1}{d}}}$.  We find
\begin{eqnarray}
K.E.&=&  \frac{ \hbar^2  }{2m} \int d^d z \frac{\rho }{N} (z) Z_{KE} \int_1^r d^d \alpha  a^{d-2}
(\rho (z) + a  \alpha^a  \partial_a  \rho + \ldots )(\frac{a}{\hbar} \alpha^a  \partial_a S + \dots )^2
\\
&=&  \frac{Z    }{2m} \int d^d z  \left [
 \rho  (\partial_a S )^2\frac{ (r^d -1) \Omega  Z_{KE} }{ N} + O(\frac{1}{N^{\frac{2}{d}}} )
\right ]
\end{eqnarray}

We now set the normalization constant to extract the kinetic energy
\f
Z_{KE} = \frac{N}{(r^d -1)\Omega }
\ff
After which we take the limit $r ^\prime $ large,  followed by $N \rightarrow \infty$.
We find the renormalized kinetic energy is 
\f
K.E. =  \int dz \rho (z) \left [
  \frac{(\partial_a S )^2   }{2m} +
O(\frac{1}{r^\prime })  +  + O(\frac{1}{N} )
\right ]
\ff

Putting this together with the above results we find
\f
H =  \int d^d z \rho (z) \left [
  \frac{(\partial_a  S)^2   }{2m} + \frac{\hbar^2}{8m} ( \frac{1}{\rho } \partial_a \rho )^2 +V + 
O(\frac{1}{r})   + O(\frac{1}{N} )
\right ]
\ff

\subsection{The symplectic measure}

The last step is to derive the continuum approximation to 
\f
S^0(w,x)  =Z_0  \int dt \sum_k p_a^k \dot{x}_k^a = - Z_0 \int dt \sum_k \dot{p}_a^k {x}_k^a
\ff
where, inspired by relative locality\cite{rl}, we integrate by parts in $dt$. The velocity of the momenta $p_a^k$  are expressed in terms of the beables as 
\f
\dot{p}_a^k=   \frac{1}{\hbar} \sum_{j \neq k}
\left [ -  \imath V^{j a}_k  \left ( \dot{S}_j -\dot{S}_k  \right ) \right ] 
\ff
The continuum approximation to this is 
\f
S^0 \rightarrow - N \int dt \int d^d z Z_0   \rho (z) z^a \dot{p}_a (z)
\ff
where
\begin{eqnarray}
 \dot{p}_a (z) &=& \int_a^r d^d x \frac{x_a}{x^2} \left ( \rho (z+x ) \dot{S}(z+x ) - \rho (z ) \dot{S}(z ) 
 \right )
 \\
 &=&  \int_a^r d^d x \frac{x_a}{x^2} x^c \partial_c \left (  \rho (z ) \dot{S}(z )  +\ldots
 \right )
 \\
  &=&  \frac{\Omega (r^d-1 )}{dN \rho (z)} \partial_a \left (  \rho (z ) \dot{S}(z ) 
 \right )
\end{eqnarray}
Consequently, with 
\f
Z_0 = \frac{dN}{\Omega (r^d-1 )}
\ff
we find
\f
S^0 =   \int dt \int d^d z \rho (z) \dot{S} (z)
\ff

Putting the three pieces together we have
\f
S =  \int dt \int d^d z \rho (z) \left [ \dot{S} +
  \frac{(\partial_a  S)^2   }{2m} +  \frac{\hbar^2}{8m} ( \frac{1}{\rho } \partial_a \rho )^2 + U + 
O(\frac{1}{r})   + O(\frac{1}{N} )
\right ]
\label{action}
\ff

\subsection{Recovery of quantum mechanics}

The action (\ref{action}) has two equations of motion which arise from varying with $\rho$ and $S$.

These are the probability conservation law 
\f
\dot{\rho} (x^a) = \partial_{a} ( \rho \frac{1}{m}  g^{ab} \partial_{b }  S(x^a)  )
\label{prob}
\ff

and the Hamilton Jacobi equation, with the addition of the quantum potential term
\f
-\dot{S}= \frac{1}{2m} g^{ab} (\frac{\partial S}{\partial x_{a} })
(\frac{\partial S}{\partial x_{b}}) + U + U^Q
\label{HJ}
\ff
where the quantum potential $U^Q$ is given by 
\f
U^Q =- \frac{\hbar^2}{2m}\frac{\nabla^2 \sqrt{\rho}}{\sqrt{\rho}}.
\ff
These are nothing but the real and imaginary parts of the Schroedinger equation, for
\f
\Psi (x,t) =\sqrt{\rho } w =  \sqrt{\rho } e^{\frac{\imath}{\hbar} S}
\ff
which we have thus shown satisfies
\f
\imath \hbar \frac{d\Psi }{dt} = \left ( - \frac{\hbar^2}{2m} \nabla^2 + U
\right ) \Psi
\ff

To complete the derivation of quantum mechanics, let us draw an important distinction between $\rho (z)$, the probability distribution for the ensemble of $N$ systems and $\rho_k (x_k)$, which is the probability distribution for a single subsystem in the ensemble.  We conjecture that over time the non-local inter-ensemble interactions (\ref{UV}) randomize the trajectories of the individual elements so that over some convergence time, $\tau$, for each $k$,
\f
\rho_k (x) \rightarrow \rho (x).
\ff

Standard arguments would suggest that this is true, but it has not been shown.  

\section{Experimental tests}

The framework proposed here is highly vulnerable to experimental test.  There are several kinds of tests possible in which departures from predictions of conventional quantum mechanics may be searched for.

\begin{itemize}

\item{}$N=1$.  Systems which have no causally indistinguishable copies in the universe are expected to behave classically, because for such systems there is no confusion possible and no quantum potential. Macroscopic systems with many incoherent degrees of freedom will be of this kind and, as already remarked on, this solves the measurement problem.  But it should be possible to use current quantum technology to engineer microscopic systems made of a modest number, $m$ of quits or electrons, which by their  combinatorial complexity cannot be expected to have any natural copies.   These would be microscopic classical systems.  It should be possible to recognize them by their spectra.

\item{}$N=2$.  If we  can engineer a single such microscopic classical systems, we can make two or several of them.  A single pair would have no quantum potential, as that is a three body interaction. 

\item{}$N=3$. At this point the quantum potential enters.   This should have consequences which are easily testable.

With $i=1,2,3$ we have, in one spatial dimension.
\f
U^{i} ( x^a_i ) = \frac{\hbar^2}{8m}  \{ \left (  \frac{1}{D(3,2)}  -   \frac{1}{D(2,1)}  \right )^2 
+  \left (  \frac{1}{D(1,3)}  -   \frac{1}{D(3,2)}  \right )^2 
+   \left (  \frac{1}{D(2,1)}  -   \frac{1}{D(1,3)}  \right )^2 \}
\ff

\item{}Modest $N$.  If we could study a sequence of experiments from $N=1$ to relatively large $N$ we could see the transition between classical behaviour for $N=1$ and quantum dynamics for large $N$.  

\item{}Non-linear  corrections to the Schroedinger equation.  We see that the Schroedinger equation receives corrections from terms such as ${\cal U}^{\Delta {\cal V}}$ in (). 
 
This leads to a modified Hamilton-Jacobi equation
\f-\dot{S}= \frac{1}{2m} g^{ab} (\frac{\partial S}{\partial x_{a \alpha} })
(\frac{\partial S}{\partial x_{b \alpha}}) + U 
- \frac{\hbar^2}{2m}\frac{\nabla^2 \sqrt{\rho}}{\sqrt{\rho}} + \Delta U^Q
\ff
where
\f
\Delta U^Q = \frac{r^{\prime 2}}{N^{\frac{2}{d}}} \frac{d}{d+2} \frac{\hbar^2}{2m}
\left [
\frac{\nabla^4 \rho }{\rho} -2 \frac{(\nabla^2 \rho )^2 }{\rho^2}
-2 \frac{(\nabla^a \rho )( \nabla_a \nabla^2 \rho  )}{\rho^2 }
\right ]
\ff
for some length scale $r'$.   This implies non-linear modifications of the Schroedinger equation
\f
\imath \hbar \frac{d\Psi }{dt} = \left ( - \frac{\hbar^2}{2m} \nabla^2 + U
 + \Delta U^Q (\bar{\Psi} \Psi ) \right ) \Psi
 \ff
 Notice that probability conservation is unaffected.
 
 To first order this perturbs energy eigenvalues
 \f
 \Delta E = \int d^d z \bar{\Psi} 
\Delta U^Q (\bar{\Psi} \Psi )  \Psi
 \ff

Such terms are very well bounded by experiment\cite{nonlinear}.

It is easy to estimate that
\f
\Delta E \approx \frac{1}{N^{\frac{2}{d}}} \frac{r^{\prime 2}}{a_0^2} \frac{\hbar^2}{2ma_0^2}
= \frac{R^{2}}{a_0^2} \frac{\hbar^2}{2ma_0^2}
\ff
so
\f 
\frac{\Delta E}{E} \approx   \frac{1}{N^{\frac{2}{d}}}  \frac{r^{\prime 2}}{a_0^2}
\ff

We learn an important lesson from this, which is that the infrared cutoff scale $R$ sets the size of the expected departures from linear quantum dynamics.  In our present treatment this is a free parameter, thus we can look for possibilities to bound it by experiment.  

\end{itemize}

\section{Comments and objections}

\subsection{Quantum statistics}

It is straightforward to derive the statistics of bosons and fermions from the assumptions enunciated above.  We require only a translation into the present notation of the standard derivations.  

Suppose the configuration space refers to $M$  identical particles in $d$ dimensions, whose positions are given by $x^a_{Ik}$ and which have phase variables $w_{k}$ (one phase for each configuration of the $M$ particles.).   Here $I=1, \ldots M$ refers to the positions of the identical particles, where as $k=1, \ldots N$ refers to the members of any additional  ensemble these are members of.

As application of the PSR we can ask why the world is as it is rather than with the positions of the first and second particles switched.
\f
x^a_{k1} \leftrightarrow  x^a_{k2},    
\ff
By saying that the particles are identical we mean that there are no differences, i.e. the system is defined for each member of the ensemble by the unordered set of positions
\f
\{ x^a_{k1},  x^a_{k2}, \ldots  \}   \equiv \{ x^a_{k2},  x^a_{k1}, \ldots  \} 
\ff
It follows that the probability densities satisfy
\f
\rho (  x^a_{k1},  x^a_{k2} , \dots , t ) = \rho (  x^a_{k2},  x^a_{k1} , \dots , t )
\ff
for all time, which implies also
\f
\dot{\rho} (  x^a_{k1},  x^a_{k2} , \dots , t ) = \dot{\rho} (  x^a_{k2},  x^a_{k1} , \dots , t )
\ff

It follows directly from (\ref{prob}) and (\ref{HJ}) that the phases satisfy 
\f
S (  x^a_{k1},  x^a_{k2} , \dots , t ) = S (  x^a_{k2},  x^a_{k1} , \dots , t ) + \phi
\ff
where $\phi$ is a constant phase.  But recalling (\ref{Sdef}) we see by doing the switch twice that $2\phi$ must be zero or a multiple of $2\pi$.  Hence we have under the switch of two particles
\f
w_k \rightarrow \pm w_k
\ff
which gives us bosonic and fermionic statistics.

It is important to make an additional point.  Consider a single system with a large number, M,  of identical particles moving in the same external potential.  These might be helium atoms in a beaker of superfluid helium or electrons in a doped semiconductor.  These constitute an ensemble of identical particles in isomorphic potentials, which happens, in this case, to be the same potential.  Hence the inter-ensemble interactions may be expected to be active here as well.  Hence, even if there are no copies of that beaker or that precise doped semiconductor in the universe, we already are dealing with large $M$'s greater than $10^{20}$.   Hence quantum mechanics may be expected to hold well in these cases.  Indeed, nothing in the derivation of quantum statistics we have just given precludes the imposition of the inter-ensemble interaction between the identical particles.   

Hence, the theory we have described here will agree with quantum mechanics for cases like these of so-called ``macroscopic" quantum systems, because they constitute an ensemble of microscopic systems all by themselves.

\subsection{Preferred simultaneities}

It may be objected that the present formulation, by virtue of its invoking interactions between distant subsystems, requires a preferred simultaneity.  Does this inhibit its application to relativistic systems?

One may note that the same criticism may be made to any completion of quantum mechanics which gives a more completed description of the trajectories of individual systems.  We know this because of a theorem of Valentini\cite{AV-signaling}.

One answer is that general relativity has recently been reformulated as a theory with preferred foliations.  Called shape dynamics\cite{sd}, this formulation trades many fingered time, or refoliation invariance, for local scale invariance.  Shape dynamics reproduces all the predictions of general relativity which have been confirmed, hence our knowledge of space-time physics is consistent with the existence of a preferred foliation.

\subsection{Defining degrees of relational similiarity}

We use informally two notions, similarity of causal pasts of events and similarity of recent pasts, or preparations, of isolated subsystems and implied they are related to each other.  The first was defined in a causal set ontology, the second within an operational framework describing subsystems.  More work needs to be done defining each of these notions and their relationship to each other.  

In this further work, the absolute notions of similar or not can be replaced by degrees of similarity.  This would be incoherent in a fundamental theory, but it is important to emphasize that the task we are engaged with is the construction of a theory of subsystems, which is necessarily approximate.  (This is related to the cosmological dilemma discussed in \cite{SU,TR,TN}.)

\section{Motivations}

The hypothesis behind this completion of quantum theory are inspired by a broad principle, which has already had enormous influence on our understanding of space and time.  This is Leibniz's Principle of sufficient reason (PSR)\cite{PSR}, which  can be stated as

\begin{itemize}

\item{} Every question of the form of {\it why is the universe like $X$ rather than $Y$} has a reason sufficient to explain why.

\end{itemize}

Big philosophical principles function in science as guides; in this spirit we may take the PSR as {\it the aspiration to eliminate arbitrary choices from the statements of the laws and initial conditions of physics.}  This aspirational version of the PSR\cite{TN,SU} has been very influential in the search for fundamental laws.  Some of the ideas it has inspired are:

\begin{itemize}

\item{}Space and time are relational. Space and time represent relational and dynamic properties that allow each subsystem of the universe to be uniquely distinguished in term of their relations to the rest of the universe\cite{relational}.  
This was the basis for the critique, made by Leibniz\cite{PSR}, Mach\cite{Mach} and others, of Newton's conception of absolute space, which inspired Einstein in the construction of its first full realization in his general relativity theory\cite{Julian-discovery}. 

\item{}  This has further consequences which we  exploited in this paper. {\it  Localization in space is a consequence of having unique relational properties, i.e. a unique causal neighbourhood.  Objects or subsystems that are hard to distinguish from similar systems should be hard to localize unambiguously, and thus may be in causal contact.}

\item{}The laws of physics have no ideal elements and depend on no fixed, non-dynamical structures\cite{Ideal}.  This is the basis of the requirement that fundamental theories be background independent, which is satisfied by general relativity in the cosmological case in which space-time is spatially compact.

\item{}Einstein's principles of causal closure and reciprocity.  Everything that influences the evolution of a subsystem of the universe is itself a part of the universe.  There are no entities which effect the evolution of degrees of freedom, which do not themselves evolve in response to influences.  {\it If an ensemble of systems influences a system then every last member of that ensemble must exist as a physical system somewhere in the universe.}

\item{}One of the most important implications of the PSR is another principle, known as the {\it principle of the identity of the indiscernible.} (PII)  This states that any two events or subsystems of the universe which are distinct have distinct properties derived from their relations with the rest of the universe\cite{PSR}.

The PII implies that fundamental, cosmological theories have no global symmetries, because a global symmetry of a cosmological theory would be a transformation between distinct states of the universe, each of which has exactly the same relational properties.  Indeed, general relativity in the cosmological case has no global symmetries or non-vanishing conservation laws\cite{karel}.

Global symmetries arise within effective theories of subsystems of the universe, and they describe transformations of a subsystem with respect to the rest of the universe which is, for purposes of the effective description, regarded as a fixed, 
non-dynamical, frame of reference\cite{TN,SU}.  

\item{}{\bf Maximal variety}   We have seen how the principle of maximal variety is realized dynamically by incorporating the negative of the variety as a potential energy.  This realizes the PII dynamically, by acting to make systems which are similar distinct.  

To illuminate the idea of variety we can describe it in the context of a causal set\cite{cs}. We start by defining the $n$'th neighbourhood of an event, $I$, called ${\cal N}_n (I)$, it consists of the subsystem consisting of all events $n$ causal links into the future or past from $I$.  Then we can call the distinguishability of two events, $I$ and $J$, ${\cal D}(I,J)= \frac{1}{n_{IJ}}$, where $n_{IJ}$ is the smallest $n$ such that ${\cal N}_n (I)$ is not isomorphic to ${\cal N}_n (J)$. 
The higher $\cal V$ is, the less effort it is to distinguish every event from every other by describing their causal neighbours.  

In \cite{variety}, Julian Barbour and I proposed two uses of the concept of the variety of a system.  First, variety measures the complexity or the organization of a system.  We showed several examples in which $\cal V$ provides an interesting measure of complexity.  For example, the variety of a city is a measure of how easy it is to know where you are by looking around.  A modern suburban development has lower variety than an old city because it has more corners from which the view is similar.  

But we also proposed a dynamical principle that the universe evolve so as to extremize its variety.  We speculated that this highly non-local dynamical principle might underlie quantum theory.  In this paper we develop this idea by showing that the quantum potential of Bohm can be understood to be a measure of the variety of a system of similar subsystems of the universe.  

\item{}The PSR also demands that we can explain how and why the particular laws which describe local physics in our universe were selected from a large set of equally consistent laws\cite{LOTC,SU,TR,TN}.  As the American philosopher Charles Sanders Peirce enunciated in the 1890's this can only be done in a way that yields falsifiable predictions if the laws are not absolute but are the result of a dynamical process of evolution\cite{Peirce}.  

\item{}We also argue in \cite{SU,TR,TN} that neither quantum nor classical mechanics mechanics can be usefully extended to a theory of the whole universe. One of several reasons is that any such extension leaves unanswered the questions of why the laws of nature and the cosmological initial conditions were chosen, thus the PSR is unfulfilled.  This means that quantum mechanics is restricted to a theory of subsystems of the universe, and it must theresor be an approximation to a cosmological theory which does not allow a free specification of laws and initial conditions.  In this paper we seek to build on this insight by constructing quantum mechanics explicitly as a theory of subsystems.

\end{itemize}

\subsection{Taking the principle of the identity of the indiscernible seriously}

Aa we seek to apply these ideas to quantum physics we should be mindful that all serious approaches to quantum gravity agree that space is emergent.  The emergence of space is a contingent property of a phase that the universe may be in.  But if space is emergent, then locality is emergent too.  This implies that how physics sorts itself out into a mix of local and non-local, in which strictly local propagation of information and energy takes place in a sea of non-local quantum correlations, must be a result of a dynamical equilibrium characterizing the low level excitations of the phase of the universe in which space emerges.

But if locality is emergent we expect defects or dis-orderings of locality\cite{dl}.  These would be represented by pairs or sets of particles which are far from each other in the classical emergent metric geometry, but which are actually nearby or adjacent in the real, fundamental causal structure.  It has been suggested before that this kind of disordered locality could be connected to quantum non-locality \cite{qmfqg}, here we propose a novel expression of this idea that connects it with the previous idea, which suggests that defects in localization should be consequence of failures to uniquely distinguish subsystems or events, in terms of their role in the dynamical network of relational properties.

Furthermore, recent work suggests that the emergent locality is relative\cite{rl}.   In past work we understood this to mean that different observers, at different places and in different states of motion ascribe different notions of locality to distant events.  Here we further relativize and radicalize the notion of locality by understanding that locality is a consequence of identity, so that only subsystems which may be uniquely identified by the network of interactions and relationships they participate in get localized uniquely in a conventional way.  

Thus,  if locality is a consequence of distinctiveness, as measured by the relations of an event or subsystem with the rest of the universe, we expect two subsystems which are very similar to each other to be near to each other, in the true microscopic causal structure.  Subsystems which are similar to each other in the sense of having nearly isomorphic relations to neighbouring events, may then be able to interact with each other, in spite of being far away from each other in the emergent spatial geometry.  In this paper we propose a form for such non-local interactions, which acts to increase the distinctiveness of pairs of subsystems.  This  is necessary to prevent violations of the PII.  We have  show in this paper that this gives rise exactly to Bohm's quantum potential, and hence to quantum mechanics.  The Schroedinger equation then emerges as a consequence of a dynamical implementation of the PII.  That principle is then re-interpreted, not as an epistemological truism, but as a dynamical principle that underlies and explains {\it ``why the quantum."}



\blankline
\blankline
We then saw that the further application of the PII to a system of identical particles then gives rise to either fermionic or bossing statistics.

It then appears that the main features of quantum physics, the statistical, indeterminate character of local laws, and their interruption by non-local correlations, are consequences of the principle of the identity of the indiscernible.  We give further arguments which support this conclusion.

\subsection{The statistical character of local physics is a consequence of the PII}

We may apply the principle of the identity of the indiscernible to show that local physics must be indeterminate on the level of fundamental systems and events.  This is an argument which appeared first in \cite{unique}.

To make this argument, we work with a causal set ontology according to which what is real is a thick present of events and processes, which create novel events from present events.  This is discussed\footnote{For a different approach to causal sets, see \cite{cohl}.} in more detail in \cite{unique,halfquantum}.   For the construction of non-relativistic quantum mechanic below, we will work, in the next sections, in a more operational framework.

By the PII, two distinct events $e$ and $f$ must have different (that is non-isomorphic) causal neighbourhoods.  The causal neighbourhood of an event, $e$, labeled $N(e)$,  is the subset of the causal set of events making up the history of the universe which 
involve $e$.  $N(e)$ is the disjoint union of a past set $P(e)$ and a future set $F(e)$.  Now it follows from the PII that if there are two distinct events $e$ and $f$ such that their past sets are isomorphic
then their future sets cannot be isomorphic.
\f
P(e) = P(f) \rightarrow  F(e) \neq F(f)
\ff
The same holds true for $e$ and $f$ any subsets of a causal set, such as a subset of an anti chain (i.e. a space like region.)  

To give this force in a large universe we can add a bit more structure.  Let 
$P(e)_n $ and  $F(e)_n$ be the causal past and future $n$ steps into the past or future.  Then we an replace () by the requirement that there exists an $n$ much less than $N_U$, the number of past events in the universe, such that
\f
P(e)_n = P(f)_n \rightarrow  F(e)_n \neq F(f)_n
\ff

In a vast universe, for fixed $n < N_U $ there are bound to be instances of $P(e)_n = P(f)_n$ holding sufficiently far into their pasts, because if interactions are simple, there will not be that many possibilities for the recent past of an elementary event. 

But, normally $F(e) \neq F(f)$ will be enforced because the causal past of $F(e)$, denoted $P[F(e)]$, is distinct from
$P[F(f)]$ because if one goes far enough in the future the pasts of the futures of the two events become distinct.  This is because, for most events, the sizes of the sets $P(e)_n $ and  $F(e)_n$ grow  like $n^d$, where $d$ is the spatial dimension.  

However there are systems that this doesn't apply to, which are {\it isolated systems.}
These are systems that have a restricted causal future that does not grow faster than linearly with $n$, as the system evolves to the future (or, similarly, into the past.).  Systems may be isolated naturally, by being sufficiently separated or shielded from their environments.  We also construct isolated systems in order to focus experiments on fundamental interactions; to isolate a system is the basic method of laboratory science.

It follows that two isolated systems $e$ and $f$ with similar environments and similar causal pasts,
$P(e) \approx P(f)$ are in danger of violating the PII.   Indeed if the laws of physics are deterministic, it is exactly those isolated systems which have $P(e) = P(f)$ which we would expect, by determinism to evolve identically such that $F(e) = F(f)$. 

So exactly how can two isolated systems avoid violating the PII?  First, if the laws of nature are statistical and indeterminate, so $P(e) = P(f)$ need not imply $F(e) = F(f)$, even if the systems are truly isolated.  This argument was developed in [].  This, I would propose is the origin of quantum statistics.

But this turns out to be not sufficient, because even if the laws are statistical, 
so that $e$ with past $P(e)$ can have several possible futures, $F(e)_I$, in a big universe with a vast number of events there still will arise by chance two distinct events $e$ and $f$ with identical causal pasts and identical causal futures.

Thus, to ensure the PII something more is needed.  This is an interaction between two isolated systems, with identical pasts that will prevent their having identical futures.  This interaction has to be repulsive (when expressed in terms of relational observables), to ensure that distinct microscopic subsystems have distinct values of their beables,  in order to prevent violations of the PII.  Here we have proposed that the variety of the ensemble be used to generate this inter-ensemble interaction.   We saw that this is the origin of the quantum potential, of deBroglie-Bohm theory.

Thus, we arrive at an ensemble interpretation of the quantum state, but the ensemble is real and not imagined; it is a real physical ensemble consisting of a finite set of similar systems which exist throughout the universe.  The ensemble does, as in dBB and Nelson, influence the individual member, but that is in accord with the principle of causal closure because that influence is just a summary of a great many multi body interactions amongst members of an isolated subsystem's ensemble.

This scheme is non-local, as we know any realist completion of quantum mechanics must be.  It is indeed wildly non-local, in order to enforce the PII in a huge universe with vast numbers of nearly identical elementary systems.  But we should not be surprised because we know from diverse studies of and approaches to quantum gravity that space and locality are expected to be emergent from a more fundamental level of description in which they play no role.  

\subsection{How the measurement problem is solved}

We see that, in a world governed by the PII, locality is a consequence of having a unique identity, and that is a property enjoyed only by systems large and complex enough that they have neither copies in the universe nor near copies.  There is in this scheme a natural definition of macroscopic: if a system is large and complex enough to have no copies (in a precise sense defined below), it is not part of any ensemble of subsystems.  It is unique on its own and hence it neither interacts with, nor an it be confused with, distant similar subsystems.  Consequently it can be stably localized.  

Such a macroscopic subsystem does not, by its uniqueness, suffer any quantum effects.  It's motion is subject only to local forces, it does not answer to any inter-ensemble interactions, hence its center of mass coordinates evolve according to the laws of classical mechanics, without a quantum potential.  That is to say, cats do not have ensembles of similar systems, and they are either dead or alive. This is then the answer to the measurement problem\cite{RE}.

\section{Conclusions}

The real ensemble hypothesis\cite{RE} has been strengthened by the use of a greatly simplified interaction between members of the ensemble of similar systems, based on the principle of extremal variety\cite{variety}.

\section*{Acknowledgements}

It is a pleasure, first of all, to thank Julian Barbour for our collaboration in the invention of the idea of maximal variety\cite{variety}, and for many years of conversations and friendship since.
This work represents a step in a research program which builds on a critique of the role of time in cosmological theories developed with Roberto Mangabeira Unger\cite{SU} and explored with Marina Cortes and, most recently Henrique Gomes.  This work develops a specific idea that emerged from that critique, which is that ensembles in quantum theory must refer to real systems, that exist somewhere in the universe\cite{RE}.    I am grateful to Lucien Hardy, Rob Spekkens and Antony Valentini for criticism of my original real ensemble formulation, as well as to  Jim Brown, Ariel Caticha, Marina Cortes, Dirk - AndrŽ Deckert, Michael Friedman, Laurent Freidel, Henrique Gomes, Michael Hall, Marco Masi, Djorje Minic, Wayne Myrvold,  John Norton, Antony Valentini and Elie Wolfe for comments on the present draft or related talks.  

This research was supported in part by Perimeter Institute for Theoretical Physics. Research at Perimeter Institute is supported by the Government of Canada through Industry Canada and by the Province of Ontario through the Ministry of Research and Innovation. This research was also partly supported by grants from NSERC, FQXi and the John Templeton Foundation.


\end{document}